\documentclass [11pt] {article}
\usepackage[utf8]{inputenc}
\usepackage[T1]{fontenc}
\usepackage{xcolor,amsmath,amssymb}
\usepackage{geometry}
\usepackage{graphicx}
\usepackage{bm}
\usepackage{bbm}
\usepackage[sort&compress,comma,square,numbers]{natbib}

\definecolor{darkblue}{rgb}{0,0,0.6}
\definecolor{darkred}{rgb}{0.6,0,0}
\usepackage[colorlinks=true,urlcolor=darkblue,citecolor=darkblue,linkcolor=darkred,
filecolor=darkred,hyperfootnotes=false]{hyperref}

\usepackage{authblk}

\usepackage{cmbright}

\renewcommand{\boldsymbol}[1]{\mathbold{#1}}

\newcommand{\ind}[1]{_{\mathrm{#1}}}

\newcommand{\FF}{\boldsymbol{F}}

\newcommand{\kk}{\boldsymbol{k}}

\newcommand{\mcT}{\mathcal{T}}
\newcommand{\uu}{\boldsymbol{u}}
\newcommand{\mcV}{\mathcal{V}}

\newcommand{\xx}{\boldsymbol{x}}

\newcommand{\eeta}{\boldsymbol{\eta}}

\newcommand{\dd}{\mathrm{d}}
\newcommand{\ed}{\mathrm{e}}
\newcommand{\id}{\mathrm{i}}

\newcommand{\transp}{^\mathrm{T}}

\begin{document}

\title{Universal Long Ranged Correlations in Driven Binary Mixtures}

\date{}



\author[1,2]{Alexis Poncet}
\affil[1]{Gulliver, CNRS, ESPCI Paris, PSL Research University, 10 rue Vauquelin, Paris, France}
\affil[2]{Département de Physique, ENS, PSL Research University, 24 Rue Lhomond, 75005 Paris, France}

\author[3]{Olivier Bénichou}
\affil[3]{Laboratoire de Physique Théorique de la Matière Condensée, CNRS/UPMC, 4 Place Jussieu, F-75005 Paris, France}

\author[1]{Vincent Démery}

\author[3]{Gleb Oshanin}





\maketitle

\begin{abstract}
\normalsize
When two populations of ``particles'' move in opposite directions, like oppositely charged colloids under an electric field or intersecting flows of pedestrians, they can move collectively, forming lanes along their direction of motion.
The nature of this ``laning transition'' is still being debated and, in
particular, the pair correlation functions, which are the key observables to quantify this phenomenon, have not been characterized yet.
Here, we determine the correlations using an analytical approach based on a linearization of the stochastic equations for the density fields, which is valid for dense systems of soft particles.
We find that the correlations decay \textit{algebraically} along the direction of motion, and have a self-similar exponential profile in the transverse direction.
Brownian dynamics simulations confirm our theoretical predictions and show that they also hold beyond the validity range of our analytical approach, pointing to a universal behavior.
\end{abstract}


\section{Introduction}\label{}

Binary driven mixtures, in which two populations of particles are driven in opposite directions, encompass a variety of systems, ranging from pedestrian traffic~\cite{Moussaid2012, Cividini2013, Bain2016} to charged particle systems like colloidal suspensions~\cite{Vissers2011b}, electrolytes~\cite{Netz2003,Demery2016b}, plasmas~\cite{Sutterlin2009}, or ionic liquids~\cite{Lee2014c}.
For a large enough driving force, such systems undergo an out-of-equilibrium transition into a state in which the particles moving in the same direction spontaneously form lanes, thereby increasing their average velocity~\cite{Helbing2000,Dzubiella2002}.
The nature of the laning transition remains unclear~\cite{Dzubiella2002, Chakrabarti2003,Chakrabarti2004,Glanz2012,Kohl2012, Klymko2016,Foulaadvand2016} and still little is known about the correlations---which are key observables to quantify the transition---and in particular about their decay at large distances~\cite{Kohl2012}.

Previous studies suggest that these correlations may be long-ranged.
Indeed, when only a small fraction of the particles is biased, the correlation between them and the unbiased particles follows the distribution of the density of particles in the wake of a \emph{single} intruder moving in a quiescent bath, which is relevant for microrheology~\cite{Meyer2006,Candelier2010,Wilson2011b,Nazockdast2016}.
This distribution has been computed for different models: Brownian dilute hard spheres~\cite{Squires2005}, dense soft spheres~\cite{Demery2014c} or hard core particles on a lattice~\cite{Benichou2000}.
In the last two cases, the density behind the intruder was found to approach its unperturbed value as a power law.

The situation is by far less clear when \emph{half} of the particles are driven in one direction. 
In this case, the correlations have only recently been investigated using numerical simulations and a numerical integration of a closure of the $N$-particles Fokker-Planck equation~\cite{Kohl2012}. 
It was shown that the correlations may decay as a power law, with an exponent between 1 and 2, and it was argued that it could be related to a diverging correlation length, and hence, to a true phase transition.
However, this observation seemingly contradicts the existence of a finite correlation length, as measured in numerical simulations of large systems~\cite{Glanz2012}.
In view of this apparent controversy, it is  desirable to have some analytical predictions on the asymptotic decay of correlations.

Here, we determine the correlations in a model driven binary mixture using an analytical approach based on a linearized Stochastic Density Functional Theory (SDFT).
This approach was introduced recently to compute the correlations in an electrolyte~\cite{Demery2016b} and is reliable for dense systems of soft particles. 
We obtain that the correlations decay {as a power law} with an exponent $(d+1)/2$ at any value of the external force, in both directions along the field, and have a universal exponential shape in the transverse direction.
Brownian dynamics simulations of harmonic spheres agree quantitatively well with our predictions, and also demonstrate that the scaling and the shape of the correlations hold beyond the validity range of our theory.
This observation suggests a universal character of the behavior of correlations.

\begin{figure}
\begin{center}
\includegraphics[width=.6\linewidth]{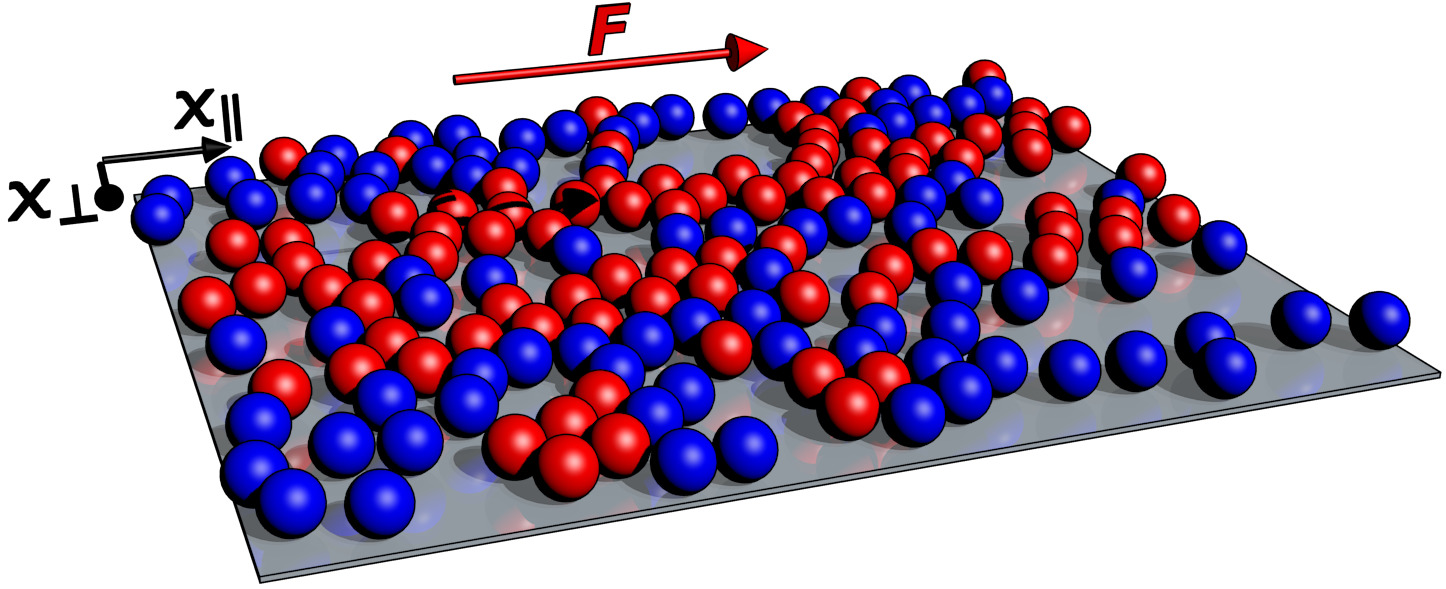}
\end{center}
\caption{{Snapshot of numerical simulations of Brownian harmonic spheres with $\bar\rho=0.5$, $\tau=0.5$, $T=10^{-4}$ and $F=0.005$}. The ``tracers'' (red) are {subject} to an external force $\FF$ while the ``non tracers'' (blue) are not. Particles of the same species tend to form lanes in the direction of the field. 
}
\label{fig:scheme}
\end{figure}

\section{Model}\label{}

We consider {a simple} model {of a} binary driven mixture which comprises  
$N$ colloidal particles, suspended in a solvent which acts as a heat bath at temperature $T$, and interacting via an isotropic pair potential $V(x)$.
A fraction $\tau=\tau_1$ of the particles---called ``the tracers'' in what follows---is also subject to an external constant force $\FF$ ~\cite{Dzubiella2002, Kohl2012,Glanz2012} (Fig.~\ref{fig:scheme}).
We assume an overdamped Langevin dynamics so that the coordinates $\xx_i(t)\in\mathbb{R}^d$ of the particles obey
\begin{equation}\label{eq:dynamics}
\dot\xx_i(t)=\mathbbm{1}_{i\in \mcT_1}\FF-\sum_{j\neq i}\nabla V(\xx_i(t)-\xx_j(t))+\eeta_i(t),
\end{equation}
where the Gaussian white noise $\eeta_i(t)$ has a correlation function $\langle \eeta_i(t)\eeta_j(t')\transp \rangle=2T\delta_{ij}\delta(t-t')\mathbbm{1}$, the mobility is set to $\kappa=1$, $\mcT_1$ is set of tracers, and $\mcT_2$ is the set of ``non-tracers'', present at fraction $1 - \tau=\tau_2$.

This model can also describe situations where the two species are submitted to different external forces, since by symmetry only the difference of the forces applied to the two species affects the correlations~\cite{Dzubiella2002}.
Following Refs.~\cite{Dzubiella2002, Kohl2012,Glanz2012}, we do not consider hydrodynamic interactions.
This approximation should be valid if the hydrodynamic radius of the particles is smaller than the range of the interaction $V(\xx)$~\cite{Khair2006}, or if the hydrodynamic interactions are screened~\cite{Long2001,Vissers2011b} (e.g., in electrophoresis).

In order to quantify the cooperative effects, we introduce the effective mobility $\kappa\ind{eff}$ of the tracers as $\langle \dot\xx_i \rangle=\kappa\ind{eff}\FF$ for $i\in\mcT_1$. 
In the absence of interactions, the mobility is $\kappa\ind{eff}=1$.
Using symmetry arguments, one can show that $\kappa\ind{eff}(\tau)=1-(1-\tau)K(\tau)$, where $K(\tau)=K(1-\tau)$~(App.~\ref{smsec:symmetry}).
The factor $1-\tau$ comes from the density of non-tracers which hinders the motion of the tracers, and $K(\tau)$ describes the cooperativity of the tracers. 

The microscopic density of the two species and the pair correlation functions are defined by
\begin{align}
\rho_\alpha(\xx,t) &=\sum_{i\in\mcT_\alpha}\delta(\xx-\xx_i(t)),\\
h_{\alpha\beta}(\xx) &=\left\langle \left[\frac{\rho_\alpha(\xx)}{\bar\rho_\alpha}-1 \right] \left[\frac{\rho_\beta(0)}{\bar\rho_\beta }-1\right] \right\rangle, \label{eq:def_correlations}
\end{align}
where $\alpha=1$ for tracers, 2 for non-tracers and $\bar\rho_\alpha=\tau_\alpha\bar\rho=\tau_\alpha N/\mcV$, $\mcV$ being the volume of the system.
Averaging Eq.~(\ref{eq:dynamics}), we find that  $K$ can be expressed {via the pair correlation functions} as~(App.~\ref{smsec:mobility_correl})
\begin{equation}\label{eq:K_pair_correlations}
K = \frac{\bar\rho}{F}\int \partial_\parallel V(\xx) h_{12}(\xx)\dd\xx;
\end{equation}
the index $\parallel$ denotes the direction parallel to $\FF$.

\section{Mean-field calculation}\label{}

In order to compute the correlation functions, we use the Dean-Kawasaki equation (or SDFT) for the density fields~\cite{Kawasaki1994,Dean1996, Demery2016b}:
\begin{equation}\label{eq:dean}
\dot\rho_\alpha=\nabla\cdot \left[T \nabla\rho_\alpha - \delta_{\alpha 1} \rho_\alpha \FF
 +\rho_\alpha \sum_\beta \nabla V \ast \rho_\beta + \rho_\alpha^{1/2} \eeta_\alpha  \right],
\end{equation}
where $\eeta_\alpha(\xx,t)$ is a Gaussian noise with correlations $\langle \eeta_\alpha(\xx,t)\eeta_\beta(\xx',t')\transp \rangle=2T\delta_{\alpha\beta}\delta(\xx-\xx')\delta(t-t')\mathbbm{1}$.
This equation is exact, but it is non-linear and contains multiplicative noise, so that it is difficult to handle it. 
However, it can be linearized by assuming small density fluctuations, i.e., $\rho_\alpha(\xx,t)-\bar\rho_\alpha=\sqrt{\bar\rho_\alpha}\phi_\alpha(\xx,t)\ll \bar\rho_\alpha$~\cite{Demery2014c}. 
Under this assumption, Eq.~(\ref{eq:dean}) reads in Fourier space (with convention $\tilde h(\kk)=\int h(\xx)\ed^{-\id\kk\cdot\xx}\dd\xx$)
\begin{align}
 \dot{\tilde\phi}_\alpha & = -k^2\sum_\beta \tilde A_{\alpha\beta}\tilde \phi_\beta+\tilde\xi_\alpha,\\
\tilde A(\kk) &= T \begin{pmatrix}
1 + \id\frac{fk_\parallel}{k^2}+\tau_1 \tilde v(\kk) & \sqrt{\tau_1\tau_2} \tilde v(\kk) \\
\sqrt{\tau_1\tau_2} \tilde v(\kk) & 1+\tau_2 \tilde v(\kk)
\end{pmatrix}
\end{align}
where $\tilde\xi(\kk,t)$ is the Gaussian noise with correlations $\langle \tilde\xi_\alpha(\kk,t)\tilde\xi_\beta(\kk',t') \rangle=2(2\pi)^d T\delta_{\alpha\beta}\delta(\kk+\kk')\delta(t-t')k^2$, $f=F/T$ is the Péclet number and $v=\bar\rho V/T$ is the rescaled potential.
The eigenvalues of $\tilde A(\kk)$ have a positive real part, hence the fluctuations of the density fields are stable.
The linearization of Eq.~\eqref{eq:dean} is valid in the large density limit, $\bar\rho\to\infty$, with $v$ kept constant, meaning that the suspension is very dense and soft~\cite{Demery2014c}.

\begin{figure}
\begin{center}
\includegraphics[scale=1]{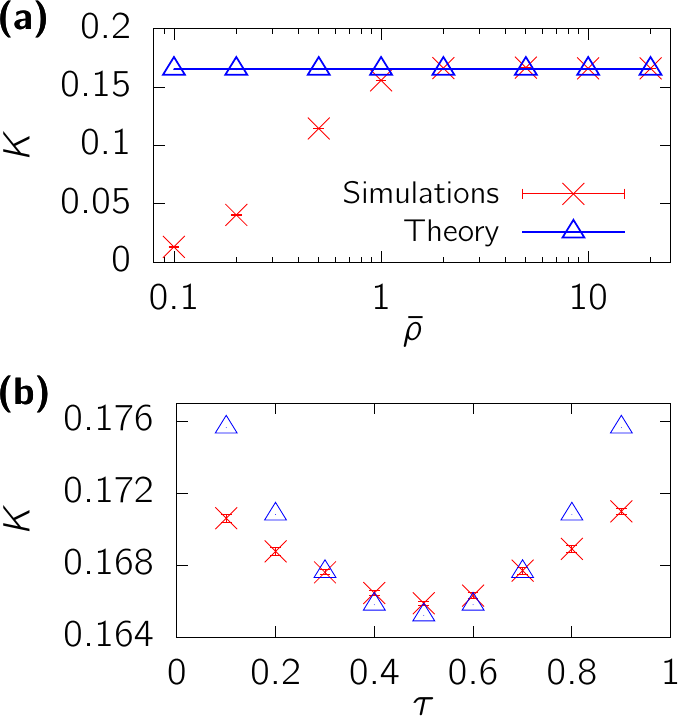}
\end{center}
\caption{{Cooperativity $K$ in $d=2$ {as a function} 
{\bf (a)}~of $\bar\rho$ with $\bar\rho/T=10$, $F/T=20$ and $\tau=0.5$, 
{\bf (b)}~of the fraction of tracers $\tau$ for $\bar\rho=2$, $\bar\rho/T=10$ and $F/T=20$. 
Solid lines and points denote analytical and simulation results, respectively.}}
\label{fig:mobility}
\end{figure}

The correlation functions of the fields $\phi$, $C_{\alpha\beta}(\xx)=\langle \phi_\alpha(\xx,t)\phi_\beta(0,t) \rangle$ are related to the pair correlation functions \eqref{eq:def_correlations} by $h_{\alpha\beta}=C_{\alpha\beta}/\sqrt{\bar\rho_\alpha\bar\rho_\beta}$.
In the stationnary regime, they satisfy~\cite{Gardiner2009}
\begin{equation}
\tilde A\tilde C+\tilde C\tilde A^*=2T\mathbbm{1}.
\end{equation}
At equilibrium, i.e., without external force, $\tilde A^*=\tilde A$, the equilibrium correlation functions $\tilde C=T\tilde A^{-1}$ are recovered.
For an arbitrary force, the solution reads~(App.~\ref{smsec:computation_correlations}):
\begin{equation}
	\tilde C = \frac{1}
	 {(1 + \tilde v) (2 + \tilde v)^2 + 
	\left(1 + \tilde v + \gamma\tilde v^2\right)
	\frac{f^2 k_\parallel^2}{k^4}} 
	\begin{pmatrix} 
	(1+\tau_2 \tilde v) \left[(2+\tilde v)^2 + \frac{f^2 k_\parallel^2}{k^4} \right] &
	-\sqrt{\gamma} \tilde v(2+\tilde v) \left(2 + \tilde v - \id \frac{fk_\parallel}{k^2}\right) \\
	-\sqrt{\gamma} \tilde v(2+\tilde v) \left(2 + \tilde v + \id \frac{fk_\parallel}{k^2}\right)  &
	(1+\tau_1 \tilde v) \left[(2+\tilde v)^2 + \frac{f^2 k_\parallel^2}{k^4} \right] 
	\end{pmatrix},
	\label{eq:sol_correl}
\end{equation}
where $\gamma=\tau(1-\tau)$.
The expression Eq. (\ref{eq:sol_correl}) allows to determine the function $K$ entering the mobility:
\begin{equation}
K=\frac{1}{\bar\rho}\int \frac{\tilde v^2(2+\tilde v)k_\parallel^2k^2}{(1+\tilde v)(2+\tilde v)^2k^4+[1+\tilde v+\gamma\tilde v^2]f^2k_\parallel^2} \frac{\dd\kk}{(2\pi)^d}.
\end{equation}
In the limit $\tau=0$, the mobility found for a single tracer is recovered (Eq. (79) of Ref.~\cite{Demery2014c}).
$K$ is minimal (i.e., the cooperativity is maximal) for $\tau=1/2$ (Fig.~\ref{fig:mobility}), consistently with the symmetry $K(\tau)=K(1-\tau)$ and the absence of cooperativity in the limit $\tau\to 0$.

\section{Large scale behavior}\label{}

The Fourier transforms of the correlation functions Eq.~\eqref{eq:sol_correl} are not continuous at $0$. 
The long distance behavior in real space can be obtained by keeping only their singular part~(App.~\ref{smsec:long_range_behavior}), leading to
\begin{equation}\label{eq:general_scaling}
h_{\alpha\beta}(\xx) \approx_{x_\parallel\to \pm\infty}  \frac{H_{\alpha\beta}^\pm}{|x_\parallel|^{\frac{d+1}{2}}} g \left( \frac{\xx_\perp}{\sqrt{D |x_\parallel|} }\right), 
\end{equation}
where
\begin{align}
g(\uu) & = \nabla_{\uu}^2 \left(\ed^{-\uu^2/2}\right)= \left(\uu^2-d+1 \right) \ed^{-\uu^2/2},\label{eq:shape}\\
D & = \frac{2(2+v_0)}{\beta f},\quad \beta=\left[1+\gamma\frac{v_0^2}{1+v_0} \right]^{1/2} \,,\\
H_{11}^\pm & = - \frac{(1-\tau) v_0^2[1+(1-\tau)v_0]\beta^{\frac{d-5}{2}}f^{\frac{d-1}{2}}}{2^{d+1} \pi^\frac{d-1}{2}\bar \rho(1+v_0)^2(2+v_0)^{\frac{d-1}{2}}}, \label{eq:H11}\\
H_{21}^\pm & = \frac{v_0\beta^{\frac{d-1}{2}}(1\mp\beta^{-1})f^{\frac{d-1}{2}}}{2^{d+1} \pi^\frac{d-1}{2} \bar\rho(1+v_0)(2+v_0)^\frac{d-1}{2}}. \label{eq:H21}
\end{align}
Eqs.~(\ref{eq:general_scaling}--\ref{eq:H21}) are our main results.
First, they not only yield the algebraic decay with power $(d+1)/2$ of the depleted wake of a single tracer~\cite{Demery2014c}, but also generalize this result by defining the shape of the wake.
At a vanishing tracers density ($\tau\to 0$), $\beta\to 1$ and  $H_{21}^+\to 0$, the decay of $h_{21}$ in front of the tracers is exponential~(App. \ref{smsec:exp_decay}).
Second, they show that for a \textit{finite} density of tracers the algebraic decay holds also for the tracer-non tracer correlation function $h_{21}$ in front of the tracers ($x_\parallel\to\infty$), and for the tracer-tracer correlation function $h_{11}$.
The tracer-tracer correlation is positive along the axis $\xx_\perp=0$, indicating a {propensity for forming} lanes.

\section{Numerical simulations}\label{}

We have performed two dimensional Brownian dynamics simulations of Eq.~\eqref{eq:dynamics} with $N=8\cdot 10^4$ harmonic spheres ($V(\xx)=(1-|\xx|)^2\theta(1-|\xx|)/2$, where $\theta(u)$ is the Heaviside function), in a square box with periodic boundary conditions~\cite{Poncet2017}~(App.~\ref{smsec:numerical_simulations}).

In Fig.~\ref{fig:mobility}a we show the cooperativity as a function of $\bar\rho$ with $\bar\rho/T$ and $f$ kept constant.
We observe that the simulation results converge to the analytical prediction as $\bar\rho\to\infty$, which is the condition of validity of our computation.
The cooperativity displays a minimum at $\tau=0.5$, which is correctly predicted by the theory (see Fig.~\ref{fig:mobility}b).

\begin{figure}
\begin{center}
	\includegraphics[scale=1]{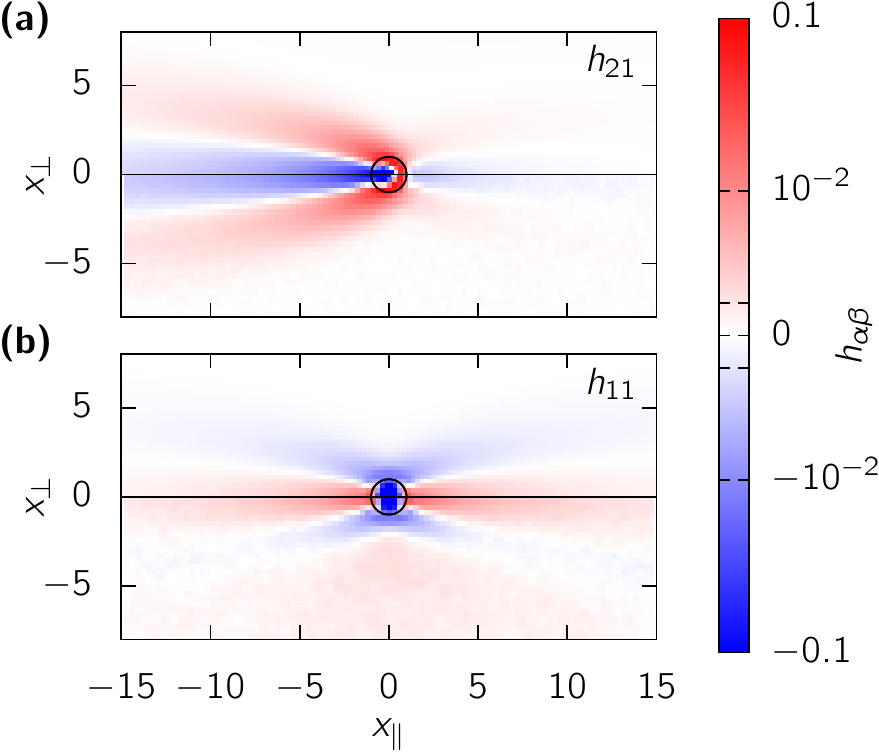}
\end{center}
	\caption{Pair correlation functions from the simulations (lower half) and the theory (upper half) for $\bar\rho = 2$, $T=0.2$, $F = 4$ and $\tau = 0.5$, in dimension $d=2$.
	The black circle {sets} the size of a particle.
	{\bf (a)}~tracer-non tracer correlation. {\bf (b)}~tracer-tracer correlation.}
	\label{fig:correl_brown}
\end{figure}

\begin{figure}
\begin{center}
	\includegraphics[scale=1]{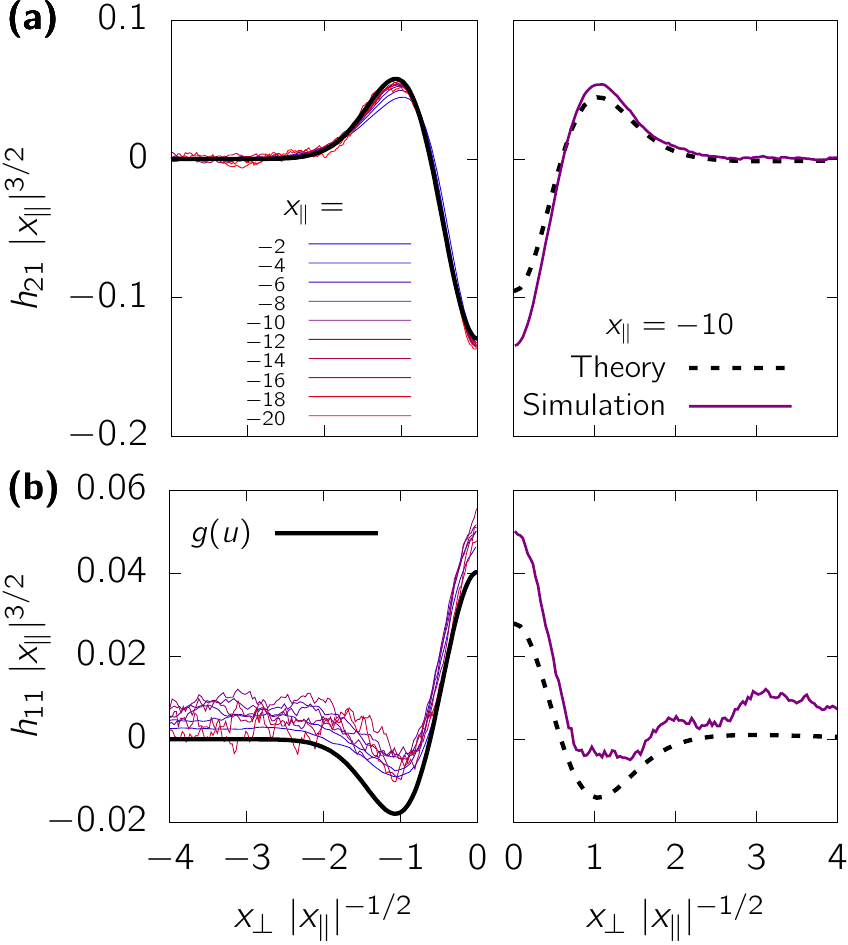}
\end{center}
	\caption{Rescaled tranverse profiles of the pair correlation functions for $\bar\rho=2$, $T=0.2$, $F = 4$, $\tau = 0.5$, in dimension $d=2$.
	\emph{Left panel:} profiles from the simulations at different longitudinal positions and universal shape Eq.~(\ref{eq:shape}) (thick black line).
	\emph{Right panel:} profiles at $x_\parallel=-10$ from the simulations (solid line) and {} the analytical prediction obtained by numerical inversion of Eq.~(\ref{eq:sol_correl}) (dashed line).
	{\bf (a)}~tracer-non tracer correlation. {\bf (b)}~tracer-tracer correlation.
	}
	\label{fig:profileBrown}
\end{figure}

The correlation functions for $\bar\rho=2$, $T=0.2$ and $f=20$ are shown in Fig.~\ref{fig:correl_brown}, and are compared to the numerical integration of Eq.~\eqref{eq:sol_correl}. 
Cuts along the transverse direction, rescaled using Eq.~\eqref{eq:general_scaling}, are compared to the numerical integration and to the shape (\ref{eq:shape}) in Fig.~\ref{fig:profileBrown}.
A very good collapse is found, which confirms the scaling laws, and a quantitative agreement is found for the shape.

\section{Universality}\label{}
\begin{figure}
\begin{center}
	\includegraphics[scale=1]{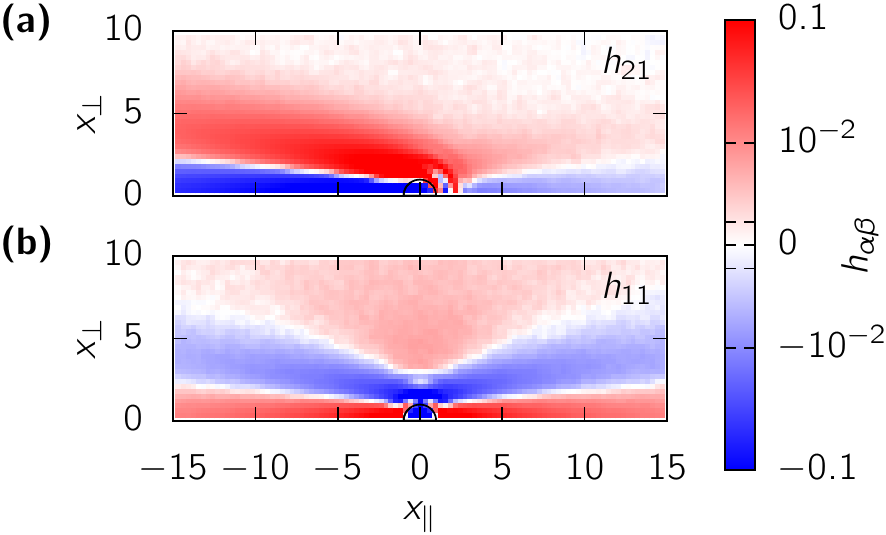}
\end{center}
	\caption{Pair correlation functions from the simulations for $\bar\rho = 0.2$,
	$T=0.001$, $F = 0.02$, $\tau = 0.5$, in dimension $d=2$. 
	The black circle {sets} the size of a particle.
	{\bf (a)}~tracer-non tracer correlation. {\bf (b)}~tracer-tracer correlation.}
	\label{fig:correlBrownHard}
\end{figure}

\begin{figure}
\begin{center}
	\includegraphics[scale=1]{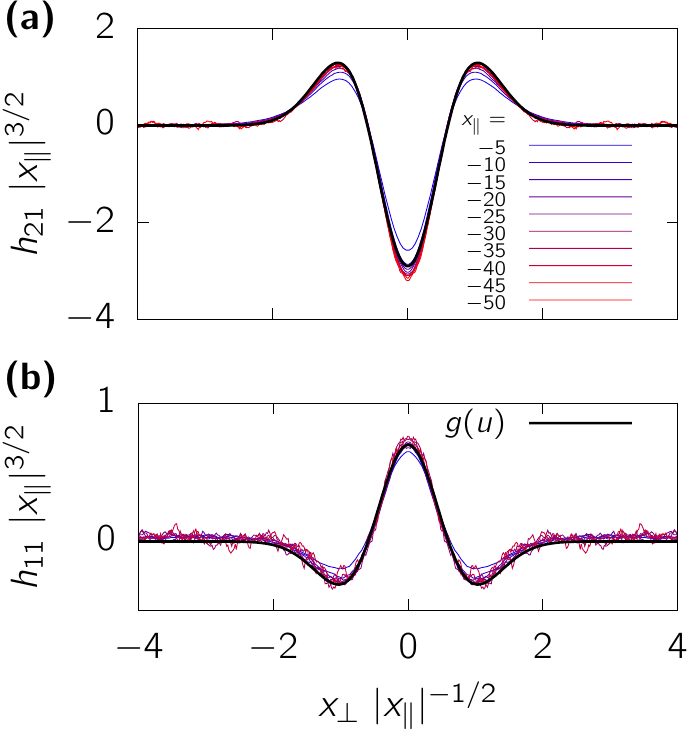}
\end{center}
	\caption{Rescaled tranverse profiles of the pair correlation functions for $\bar\rho = 0.2$,
	$T=0.001$, $F = 0.02$, $\tau = 0.5$, in dimension $d=2$.
	The transverse profiles from the simulations at different longitudinal positions are shown together with the universal shape Eq.~(\ref{eq:shape}) with fitted height and width (thick black line).
	{\bf (a)}~tracer-non tracers correlation. {\bf (b)}~tracer-tracer correlation.}
	\label{fig:profileBrownHard}
\end{figure}

We now show that the scaling~\eqref{eq:general_scaling} of the correlations and their shape \eqref{eq:shape}, hold far beyond the validity regime of our computation, namely a dense and soft suspension.
We simulated our system of harmonic spheres in the dilute and hard regime: $\bar\rho=0.2$, $T=10^{-3}$, $f=20$.
The correlations are shown in Fig.~\ref{fig:correlBrownHard}: they have the same shape as in the opposite limit of dense and soft particles. 
Rescaled transverse cuts are shown on Fig.~\ref{fig:profileBrownHard}: a very good collapse is found, and the shape is well described by Eq.~(\ref{eq:shape}) with fitted height and width.

These results suggest that the long-range behavior of the correlations, Eqs.~(\ref{eq:general_scaling},\ref{eq:shape}), is universal.
It has a simple interpretation as a diffusive process.
Replacing $x_\parallel$ (or $-x_\parallel$, depending on the direction of interest) by the time and $\xx_\perp$ by the position, Eq.~\eqref{eq:general_scaling} is the solution of a diffusion equation where the source is the second derivative of a Dirac distribution (notably, it is symmetric and its integral is zero).
This behavior emanates from simple properties of our system: 
(i) The long-distance dynamics without external force is diffusive (i.e., the relaxation rate of $\tilde\phi(\kk)$ scales as $k^2$ for small $k$);
(ii) The tracers have a finite average velocity;
(iii) The number of particles is conserved.
We conclude that the large scale behavior of the correlations is universal and should be observed in the systems sharing these properties.

\section{Discussion}\label{}

We have shown that the correlation functions in a simple model of a driven binary mixture decay algebraically in the direction of the external force, and have a self-similar profile in the direction perpendicular to the external force.
This shape is universal and should hold in any system which is diffusive at large scales and where the external field induces a finite average velocity of the particles.

Our results shed light on the previous observations.
First, we confirm the conjecture of Ref.~\cite{Kohl2012}: the correlation functions are long-ranged with an exponent $3/2$ in $d=2$, which is compatible with the value reported in Ref.~\cite{Kohl2012}.
However, our conclusion is different: (i) this algebraic decay is not associated with the divergence in the structure factors, (ii) it does not indicate a phase transition, instead it is observed over a whole region of the parameter space.
Second, in agreement with Ref.~\cite{Glanz2012}, we find no trace of a phase transition.
However, we show that the correlation length defined in Ref.~\cite{Glanz2012} does not correspond to an exponential decay of the correlations, as is usually the case; 
resolving this apparent contradiction certainly deserves future work.

In the model considered here, the system is fluid, the dynamics is overdamped and the interactions are short-ranged.
However, driven binary mixtures may be very different: they may have an underdamped dynamics~\cite{Liu2008,Sutterlin2009,Demery2015} (e.g. in plasmas), or even an active dynamics~\cite{Bain2016}, long-ranged interactions (electrostatic~\cite{Demery2016b} or hydrodynamic), a solid-like behavior in the absence of an external force (e.g., if the system is glassy)~\cite{Gazuz2009}, etc. 
It seems natural to inquire next how these differences would affect the large distance behavior of the correlation functions, and to relate it with the nature of the laning transition in these systems.


\section*{Acknowledgments}\label{}
\addcontentsline{toc}{section}{Acknowledgments}

The authors thank D. S. Dean for stimulating discussions.

\appendix

\section{Symmetry arguments for the effective mobility}\label{smsec:symmetry}

Here we link the effective mobility at a tracer fraction $1-\tau$ to the one at a fraction $\tau$.
Let $F_{/1}$ (resp. $F_{/2}$) the force applied to one tracer (resp. one non-tracer)
by all the other particles, the effective mobility is
\begin{equation}
\kappa\ind{eff}(\tau)=\frac{\langle \dot x_{1\parallel} \rangle}{F}=\frac{F+\langle F_{/1} \rangle}{F}=1+\frac{\langle F_{/1} \rangle}{F}.
\end{equation}

Let us now add a force $-F$ on all the particles. 
This leads to consider the particles 2 as the tracers.
But the addition of an external force only translates the system at a constant velocity $-F$, hence the correlation functions are not modified, and neither do the internal forces ($F_{/1}$ and $F_{/2}$).
The effective mobility as tracer fraction $1-\tau$ is given by
\begin{equation}
\kappa\ind{eff}(1-\tau)=\frac{\langle \dot x_{2\parallel} \rangle}{-F}=\frac{-F+\langle F_{/2} \rangle}{-F}=1-\frac{\langle F_{/2} \rangle}{F}.
\end{equation}
Introducing $\Delta\kappa=1-\kappa\ind{eff}$, we get
\begin{equation}
\frac{\Delta\kappa(\tau)}{\Delta\kappa(1-\tau)} = - \frac{\langle F_{/1}\rangle}{\langle F_{/2}\rangle}.
\end{equation}

The third law of Newton for the inter-particle interactions gives:
\begin{equation}
 \tau \langle F_{/1} \rangle + (1-\tau) \langle F_{/2} \rangle = 0,
\end{equation}
leading to
\begin{equation}
 \frac{\Delta\kappa(\tau)}{\Delta\kappa(1-\tau)} = \frac{1-\tau}{\tau}.
\end{equation}
Introducing $K(\tau)=\Delta\kappa(\tau)/(1-\tau)$ allows to rewrite it as
\begin{align}
 \kappa_\text{eff}(\tau) &= 1 - (1-\tau) K(\tau),\\
 K(\tau) & = K(1-\tau).
\end{align}

\section{Effective Mobility From The Correlation Functions} \label{smsec:mobility_correl}

In this section, we prove Eq.~(\ref{eq:K_pair_correlations}).
Averaging Eq.~(\ref{eq:dynamics}) and summing over the tracers, we get
\begin{equation}
\sum_{i\in\mcT}\langle \dot\xx_i \rangle=\tau N\FF-\sum_{i\in\mcT}\sum_{j\neq i} \langle \nabla V(\xx_i-\xx_j) \rangle.
\end{equation}
The left-hand side is $\tau N\kappa\ind{eff}\FF$, hence a projection onto the direction parallel to the force leads to
\begin{equation}\label{eq:keff_correlations}
\kappa\ind{eff}=1-\frac{1}{\tau N F} \left\langle \sum_{i\in\mcT}\sum_{j\neq i} \partial_\parallel V(\xx_i-\xx_j) \right\rangle.
\end{equation}
Since the potential is isotropic, $\nabla V(0)=0$; this allows us to include all the particles in the second sum in the previous equation.
The sums can then be rewritten using the density fields $\rho_\alpha(\xx)$:
\begin{equation}
\sum_{i\in\mcT}\sum_{j} \partial_\parallel V(\xx_i-\xx_j)=\int \rho_1(\xx)[\rho_1(\xx')+\rho_2(\xx')] \partial_\parallel V(\xx-\xx')\dd \xx\dd\xx'.
\end{equation}

The average can be expressed with the pair correlation functions defined in Eq.~(4), $\langle \rho_\alpha(\xx)\rho_\beta(\xx') \rangle=\bar\rho_\alpha\bar\rho_\beta[1+h_{\alpha\beta}(\xx-\xx')]$.
This leads to
\begin{align}
\left\langle \sum_{i\in\mcT}\sum_{j} \partial_\parallel V(\xx_i-\xx_j) \right\rangle & = \mcV\bar\rho_1^2\int[1+h_{11}(\xx)]\partial_\parallel V(\xx)\dd\xx + \mcV\bar\rho_1\bar\rho_2\int[1+h_{12}(\xx)]\partial_\parallel V(\xx)\dd\xx\\
& = \mcV\bar\rho_1\bar\rho_2 \int h_{12}(\xx)\partial_\parallel V(\xx)\dd\xx.
\end{align}
The last equation is obtained from the fact that $1$ and $h_{11}(\xx)$ are even functions of $\xx$, so that their product with $\partial_\parallel V(\xx)$, which is an odd function of $\xx$, has a vanishing integral.

Introducing this result in Eq.~(\ref{eq:keff_correlations}), we get
\begin{equation}
\kappa\ind{eff}=1-\frac{\bar\rho_2}{F}\int h_{12}(\xx)\partial_\parallel V(\xx)\dd\xx.
\end{equation}
Finally, using $\bar\rho_2=(1-\tau)\bar\rho$ and the definition of $K$, leads to Eq.~(\ref{eq:K_pair_correlations}):
\begin{equation}
K = \frac{\bar\rho}{F}\int \partial_\parallel V(\xx) h_{12}(\xx)\dd\xx.
\end{equation}

\section{Computation of the correlation functions}\label{smsec:computation_correlations}

Our goal is to solve
\begin{equation}
 \tilde A \tilde C + \tilde C \tilde A^\ast  = 2T \mathbbm{1}
 \label{eq:appCorrelEq}
\end{equation}

We denote
\begin{align}
 \tilde A &= T \begin{pmatrix}\alpha + \id \delta & \beta \\ \beta & \gamma \end{pmatrix},\\
 \tilde C & = \begin{pmatrix}c_{11} & c_{12} \\ c_{21} & c_{22} \end{pmatrix},
\end{align}
where
\begin{align} \label{eq:appDefParams}
\alpha & = 1+\tau_1 \tilde v,\\
\beta & = (\tau_1\tau_2)^{1/2}\tilde v,\\
\gamma & = 1+\tau_2\tilde v\\
\delta & = \frac{fk_\parallel}{k^2},
\end{align}
and $c_{11}$, $c_{12}$, $c_{21}$ and $c_{22}$ are the unknowns.

Our system of 4 equations with four unknowns (\ref{eq:appCorrelEq}) writes:
\begin{multline}
	\left\{
		\begin{array}{rl}
	 (\alpha+\id\delta) c_{11} + \beta c_{21}  +  c_{11} (\alpha -\id\delta) +c_{12} \beta & = 2 \\
	 (\alpha+\id\delta) c_{12} + \beta c_{22}  +  c_{11} \beta + c_{12} \gamma & = 0 \\
	 \beta c_{11} + \gamma c_{21}  +  c_{21} (\alpha -\id\delta) +c_{22} \beta & = 0 \\
	 \beta c_{12} + \gamma c_{22}  +  c_{21} \beta +c_{22} \gamma & = 2
		\end{array}
	\right.
	\Leftrightarrow
	\left\{
		\begin{array}{rl}
			2\alpha c_{11} + \beta (c_{12} + c_{21}) & = 2 \\
			\beta (c_{11} + c_{22}) + (\alpha + \gamma + \id\delta) c_{12} & = 0 \\ 
			\beta (c_{11} + c_{22}) + (\alpha + \gamma - \id\delta) c_{21} & = 0 \\ 
			2\gamma c_{22} + \beta (c_{12} + c_{21}) & = 2
		\end{array}
	\right. \\
	\Leftrightarrow
	\left\{
		\begin{array}{l}
			\alpha c_{11} = \gamma c_{22} \\
			(\alpha + \gamma + \id\delta) c_{12} = (\alpha + \gamma - \id\delta) c_{21} \\ 
			\beta (\frac{\gamma}{\alpha} + 1) c_{22} + (\alpha + \gamma - \id\delta) c_{21} = 0 \\ 
			2\gamma c_{22} + \beta (\frac{\alpha+\gamma-\id\delta}{\alpha+\gamma+\id\delta} + 1) c_{21} = 2
		\end{array}
	\right.
	\Leftrightarrow
	\left\{
		\begin{array}{l}
			c_{11} = \frac{\gamma}{\alpha} c_{22} \\
			c_{12} = \frac{\alpha+\gamma-\id\delta}{\alpha+\gamma+\id\delta}c_{21} \\ 
			c_{21} = -\frac{1}{\alpha+\gamma-\id\delta}\frac{\beta(\alpha+\gamma)}{\alpha} c_{22} \\
			\left[ 2\gamma - 
			\frac{\beta}{\alpha+\gamma -i \delta} (\frac{\alpha+\gamma-\id\delta}{\alpha+\gamma+\id\delta} +1)
			\beta (\frac{\gamma}{\alpha} + 1) \right] c_{22} = 2
		\end{array}
	\right.
\end{multline}
We can solve for $c_{22}$:
\begin{equation}
c_{22} = \frac{\alpha \left[ (\alpha+\gamma)^2 +\delta^2 \right] }
	{\alpha\gamma \left[ (\alpha+\gamma)^2 +\delta^2 \right] -\beta^2 (\alpha+\gamma)^2}.
\end{equation}
And the rest follows,
\begin{equation}
 \tilde C_\Phi = 
 \frac{1}{(\alpha+\gamma)^2 (\alpha\gamma - \beta^2) +\alpha\gamma\delta^2}
 \begin{pmatrix}
 \gamma \left[(\alpha+\gamma)^2 + \delta^2\right] &
 -\beta(\alpha+\gamma)(\alpha+\gamma - \id\delta)  \\
 -\beta(\alpha+\gamma)(\alpha + \gamma + \id\delta)  &
 \alpha \left[(\alpha+\gamma)^2 + \delta^2\right]
 \end{pmatrix}
\end{equation}

Finally, from the values of $\alpha$, $\beta$, $\gamma$ and $\delta$ (\ref{eq:appDefParams}):
\begin{multline}\label{eq:result_correl}
	\tilde C = 
	\left[ (1 + \tilde v) (2 + \tilde v)^2 + 
	\left(1 + \tilde v + \tau_1\tau_2\tilde v^2\right)
	\frac{f^2 k_\parallel^2}{k^4} \right]^{-1} \\
	\begin{pmatrix}
	(1+\tau_2\tilde v) \left[(2+\tilde v)^2 + \frac{f^2 k_\parallel^2}{k^4} \right] &
	-\sqrt{\tau_1\tau_2} \tilde v(2+\tilde v) \left(2 + \tilde v - \id \frac{fk_\parallel}{k^2}\right) \\
	-\sqrt{\tau_1\tau_2} \tilde v(2+\tilde v) \left(2 + \tilde v + \id \frac{fk_\parallel}{k^2}\right)  &
	(1+\tau_1\tilde v) \left[(2+\tilde v)^2 + \frac{f^2 k_\parallel^2}{k^4} \right]
	\end{pmatrix}
\end{multline}

\section{Long-range behavior of the correlation functions}\label{smsec:long_range_behavior}

\subsection{Singular part of the Fourier transforms of the correlation functions}\label{}

The behavior of a function at large distances is related to the regularity of its Fourier transform.
The fact that the Fourier transform of the correlation function is not continuous at $\kk=0$ (the limits $k_\parallel\to 0$ and $\kk_\perp\to 0$ do not commute) points to an algebraic decay in real space.
In order to determine the long range properties of the correlation function, we can extract the singular part $\tilde C\ind{s}(\kk)$ of its Fourier transform $\tilde C(\kk)$ (Eq.~(\ref{eq:sol_correl})) and compute its inverse Fourier transform.
The singular part is defined by the fact that $\tilde C(\kk)-\tilde C\ind{s}(\kk)$ is more regular than $\tilde C(\kk)$ at $\kk=0$, hence the inverse Fourier transform of this difference decays faster than the correlation function in real space. 
With this definition, the singular part $\tilde C\ind{s}(\kk)$ is defined up to a regular function.

First, the potential $\tilde v(k)$ is regular at 0, we can thus replace $\tilde v(\kk)$ by $\tilde v(0)=v_0$ in the singular part. This means that at large distances, the interaction potential acts effectively as a Dirac distribution.
The terms responsible for the discontinuity are those of the form $k_\parallel/k^2$.
Multiplying the numerator and denominator of the correlation function by $k^4$, we see that the singular part can be obtained by replacing $k^2=k_\parallel^2+k_\perp^2$ by $k_\perp^2$.
We thus obtain for the singular part:
\begin{multline}
	\tilde C\ind{s}(\kk)=
	\left[ (1 + v_0) (2 + v_0)^2 + 
	(1 + v_0 + \tau_1\tau_2v_0^2)
	\frac{f^2 k_\parallel^2}{k_\perp^4} \right]^{-1} \\
	\begin{pmatrix} 
	(1+\tau_2 v_0) \left[(2+v_0)^2 + \frac{f^2 k_\parallel^2}{k_\perp^4} \right] &
	-\sqrt{\tau_1\tau_2} v_0(2+v_0) (2 + v_0 - \id \frac{fk_\parallel}{k_\perp^2}) \\
	-\sqrt{\tau_1\tau_2} v_0(2+v_0) (2 + v_0 + \id \frac{fk_\parallel}{k_\perp^2})  &
	(1+\tau_1 v_0) \left[(2+v_0)^2 + \frac{f^2 k_\parallel^2}{k_\perp^4} \right]
	\end{pmatrix}
	\label{eq:sol_correl}
\end{multline}
As we see in the next subsection, the inverse Fourier transform of this function can be computed analytically.

\subsection{Inverse Fourier transform of the singular part}\label{}

We compute the inverse Fourier transform of the singular part of the Fourier transform of the correlation function, $\tilde C\ind{s}(\kk)$.
In the remainder of this section, we drop the subscript ``s'' in $\tilde C\ind{s}$ for brevity.

We are interested in $\tilde C_{21}$ and $\tilde C_{11}$ that we write
\begin{align}
\tilde C_{21}(\kk)&= A \frac{k_\perp^2(\alpha k_\perp^2+\id fk_\parallel)}{\alpha^2k_\perp^4+\beta^2 f^2 k_\parallel^2},\label{eq:c21_long}\\
\tilde C_{11}(\kk)&= B \frac{\alpha^2 k_\perp^4+ f^2 k_\parallel^2}{\alpha^2 k_\perp^4+\beta^2f^2k_\parallel^2},
\end{align}
where
\begin{align}
A & = -[\tau(1-\tau)]^{1/2}\frac{v_0(2+v_0)}{1+v_0}, \label{eq:def_A}\\
B & = \frac{1+(1-\tau)v_0}{1+v_0},\\
\alpha & = 2+v_0, \label{eq:def_alpha}\\
\beta & = \left(1+\frac{\tau(1-\tau)v_0^2}{1+v_0} \right)^{1/2}. \label{eq:def_beta}
\end{align}

First, we compute the inverse Fourier transform along the direction of the force,
\begin{equation}
C(x_\parallel,\kk_\perp)=\int_{-\infty}^\infty \ed^{\id x_\parallel k_\parallel}\tilde C(k_\parallel,\kk_\perp)\frac{\dd k_\parallel}{2\pi}.
\end{equation}
We compute this integral using the residue theorem, leading to
\begin{align}
C_{21}(x_\parallel,\kk_\perp) &= \frac{A}{2\beta f}\left[(1+\beta^{-1})\theta(-x_\parallel)+ (1-\beta^{-1})\theta(x_\parallel)\right] \nonumber\\
&\qquad\times k_\perp^2\ed^{-\frac{\alpha|x_\parallel|}{\beta f} k_\perp^2},\\
C_{11}(x_\parallel,\kk_\perp) &= \frac{B \alpha(1-\beta^{-2})}{2\beta f}k_\perp^2\ed^{-\frac{\alpha|x_\parallel|}{\beta f}k_\perp^2}+\frac{B}{\beta^2}\delta(x_\parallel).
\end{align}
The Dirac term in the tracer-tracer correlation function is the contribution of a tracer with itself. It has no effect on the long-distance behavior.
The inverse Fourier transform in the transverse direction is straightforward to compute: we recognize the second derivative of a Gaussian. We get Eqs.~(\ref{eq:general_scaling}--\ref{eq:H21}).

\section{Tracer-non tracer correlation in front of the tracer at vanishing tracer fraction}\label{smsec:exp_decay}

We focus on the behavior of the tracer-non tracer correlation $C_{21}(\xx)$ for $x_\parallel>0$, in the limit of vanishing tracer fraction, $\tau\to 0$.
Since we are interested in distances large compared to the range of the potential, we may approximate it by a Dirac: $v(\xx)\to v_0\delta(\xx)$, where $v_0=\tilde v(0)=\int v(\xx)\dd\xx$.
Thus, we want to inverse Fourier transform
\begin{equation}
\tilde C_{21}(\kk)= A \frac{k^2}{\alpha k^2-\id f k_\parallel},
\end{equation}
where $A$ and $\alpha$ are given by Eqs.~(\ref{eq:def_A}) and (\ref{eq:def_alpha}).
Using $k^2=k_\parallel^2+k_\perp^2$, we can compute the inverse Fourier transform along $k_\parallel$ with the residue theorem, leading to:
\begin{equation}
C_{21}(x_\parallel,\kk_\perp)=-\frac{A}{2\alpha q}\left(q+\frac{f}{2\alpha} \right)^2\ed^{-x_\parallel \left(q+\frac{f}{2\alpha} \right)},
\end{equation}
where we have introduced $q=\left(k_\perp^2+\frac{f^2}{4\alpha^2} \right)^{1/2}$.
Since we are interested in the behavior at large $x_\parallel$, we can use the saddle-point method to compute the inverse Fourier transform in the transverse direction, leading to
\begin{equation}
C_{21}(\xx)\underset{x_\parallel\to\infty}{\sim} -\frac{Af^{\frac{d+1}{2}}}{(2\pi)^\frac{d-1}{2}\alpha^\frac{d+3}{2}x_\parallel^\frac{d-1}{2}}\ed^{-\frac{fx_\parallel}{\alpha}-\frac{fx_\perp^2}{4\alpha x_\parallel}}.
\end{equation}
The first term in the exponential shows that the decay is exponential.
We remark that $A$ is negative, hence the correlation is positive: there is a finite size ``traffic jam'' in front of the tracers.
We also note that the factor $[\tau(1-\tau)]^{1/2}$ in $A$, which goes to zero when $\tau\to 0$, is removed if we consider the pair correlation function $h_{21}(\xx)$.

\section{Numerical simulations}\label{smsec:numerical_simulations}

\subsection{Details of the simulations}\label{}

We have run our simulations in two different regimes.
In the regime of a dense suspension of soft particles (Figs. 3 and 4), we used $\rho=2$, $T=0.2$, $F=4$, $\tau=0.5$, and $N=8\cdot 10^4$.
We simulated Eq.~(1) using a time step $\delta t=10^{-4}$. 
Initially, the particles are randomly placed in the square box, and the simulations was run during $2\times 10^4$ iterations (corresponding to a time $t\ind{transient}=2$) without performing any measurement in order for the system to reach its stationnary state, and then quantities where measured for $10^6$ iterations (i.e., during a time $t\ind{ss}=100$). 
The correlation functions have been measured with a resolution $\delta x=0.1$.

\begin{figure}
\begin{center}
\includegraphics[]{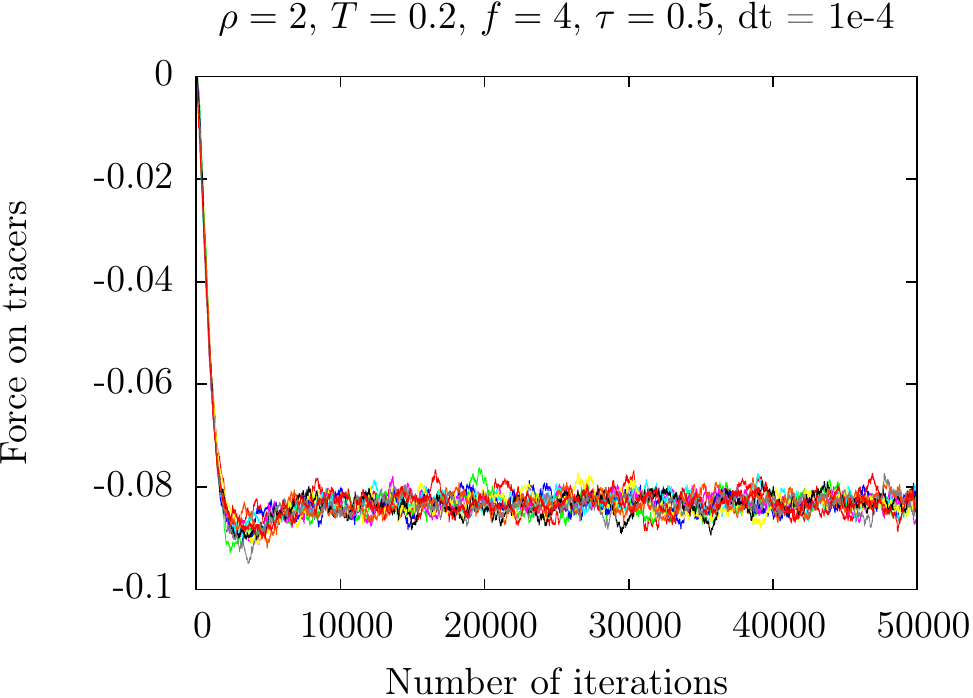}
\end{center}
\caption{Average force on the tracers in the direction of $\FF$ as a function of the number of iterations from the initial configurations where the particles are placed at random, for $\rho=2$, $T=0.2$, $F=4$, $\tau=0.5$, and $N=8\cdot 10^4$, for $10$ different simulations.}
\label{fig:steady_state}
\end{figure}

The convergence of the average force on the tracers starting from the initial random configuration is shown in Fig.~\ref{fig:steady_state} for 10 different simulations); it shows that the stationnary state is reached after $2\times 10^4$ iterations.
For each set of parameters, the observables are then averaged over a set of 50 simulations.

In the regime of a dilute suspension of hard particles (Figs. 5 and 6), we used $\rho=0.2$, $T=0.001$, $F=0.02$, $\tau=0.5$, $\delta t=0.1$, $N=8\cdot 10^4$ and the same number of iterations and independent simulations.

\subsection{Code}\label{}

The code used in our simulations is released as a free software under CeCILL 2.1 license~\cite{Poncet2017}.





\subsection{Finite size effects}\label{}

In order to check that finite size effects are absent in our numerical simulations, we have run the simulations for the parameters used in Figs.~\ref{fig:correl_brown} and~\ref{fig:profileBrown}, for different system sizes: $N\in\{1,2,5,10,20,40,80\}\times 10^3$.

\begin{figure}
\begin{center}
\includegraphics[]{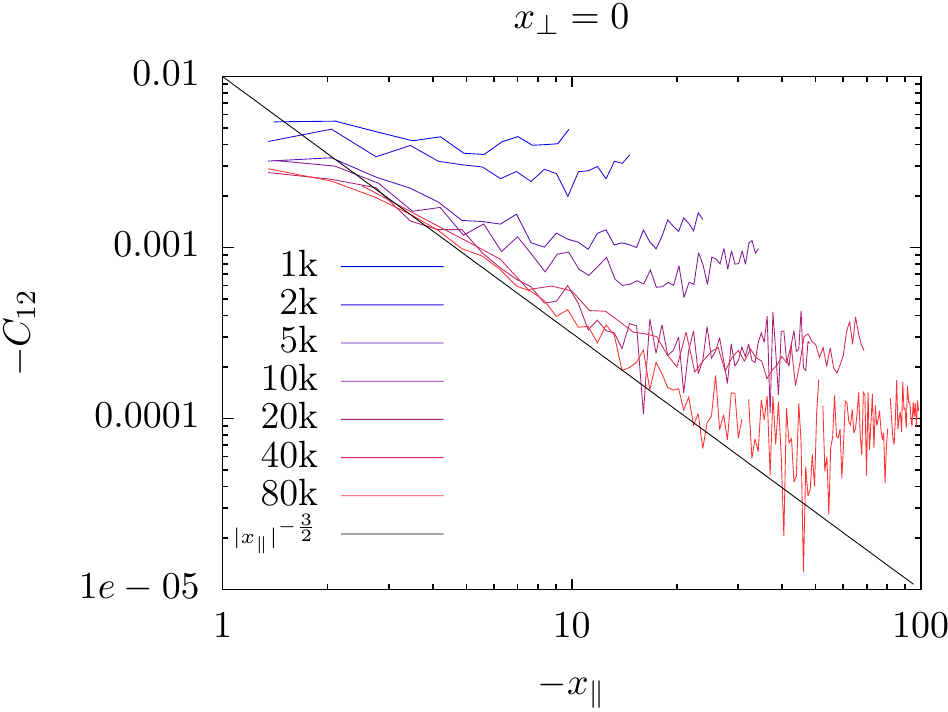}
\end{center}
\caption{Correlation function $C_{12}$ as a function of $x_\parallel$ for $\xx_\perp=0$ and different system sizes, for $\bar\rho = 2$, $T=0.2$, $F = 4$ and $\tau = 0.5$.}
\label{fig:finite_size_decay}
\end{figure}

\begin{figure}
\begin{center}
\includegraphics[]{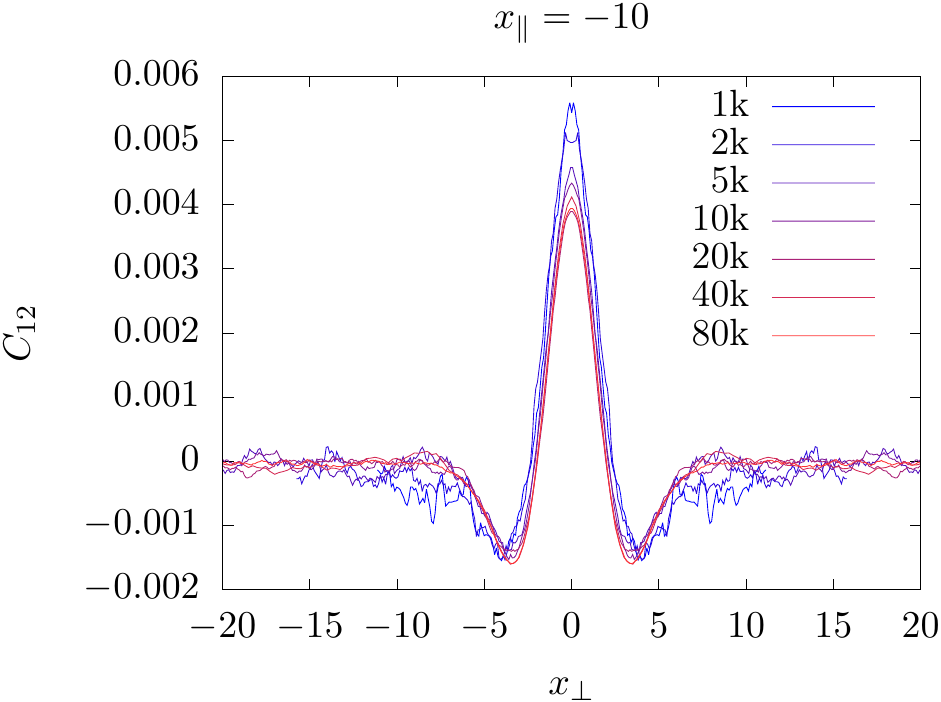}
\end{center}
\caption{Transverse profile of the correlation $C_{12}$ at $x_\parallel=-10$ for different system sizes and $\bar\rho = 2$, $T=0.2$, $F = 4$ and $\tau = 0.5$.}
\label{fig:finite_size_profile}
\end{figure}

In Fig.~\ref{fig:finite_size_decay}, we show the decay of the correlation function $C_{12}$ in the direction of $\FF$, i.e., $\xx_\perp=0$.
Finite size effects are indeed present, but they show up at larger and larger distances as the system size increases. 
In Fig.~\ref{fig:finite_size_profile}, the transverse profile of the correlation function $C_{12}$ is shown for $x_\parallel=10$; it depends on the size of the system but it converges as the size increases.

In the dilute and hard regime (Figs.~\ref{fig:correlBrownHard} and~\ref{fig:profileBrownHard}), we can use the results of Glanz and Löwen~\cite{Glanz2012}, who used similar parameters (albeit with a different interaction potential $V(x)$). They found that for $N\simeq 8\cdot 10^4$, finite size effects may be expected (i.e., the correlation length becomes comparable to the size of the system) for Péclet numbers $f\gtrsim 120$, which is well above the Péclet number that we used, $f=20$.

\end{document}